\documentclass[12pt]{iopart}

\usepackage{upgreek}
\usepackage{array}
\usepackage[dvips]{graphicx}
\usepackage{color}
\begin{document}

\title{Structural investigation of LaAlO$_3$ up to 63 GPa}

\author{Mael Guennou$^1$, Pierre Bouvier$^{1,2}$, Gaston Garbarino$^{2}$, Jens Kreisel$^1$}
\address{$^1$Laboratoire des Mat\'eriaux et du G\'enie Physique, CNRS, Grenoble Institute of Technology, MINATEC,
3 parvis Louis N\'eel, 38016 Grenoble, France}
\address{$^2$European Synchrotron Radiation Facility (ESRF), BP 220, 6 Rue Jules Horowitz, 38043 Grenoble Cedex, France}
\eads{\mailto{mael.guennou@supaero.org}, \mailto{pierre.bouvier@grenoble-inp.fr}}
\begin{abstract}
We report a high-pressure single-crystal synchrotron x-ray diffraction on a LaAlO$_3$ single crystal. The transition from rhombohedral to cubic at 14.8~GPa is confirmed by the loss of the superstructure reflections, whose intensity shows a linear pressure dependence, characteristic of a second-order transition. The crystal remains cubic up to 63 GPa, the highest pressure reached, which provides a confirmation over a very large pressure range of the general rules for the evolution of distortions of perovskites under pressure. We give the parameters of Birch-Murnaghan equations of state in the low- and high-pressure phases and discuss the evolution of the bulk modulus. \end{abstract}

\submitto{\JPCM}
%\maketitle

\section{Introduction}

In the past, much progress in the understanding of properties of ferroic $AB$O$_3$ perovskites, such as ferroelasticity or ferroelectricity, and their related phase transitions has been achieved through temperature-, or chemical composition-dependent investigations. The use of pressure has been much rarer and was mostly limited to pressures below 10 GPa due to experimental difficulties which have now been overcome for a number of years by the use of diamond-anvil cells. One of the attracting properties of the external parameter high-pressure is its character of a "cleaner" variable, since it acts only on interatomic distances \cite{Samara2000}. Furthermore, external pressure leads to otherwise unachievable reductions of volume and chemical bond lengths. Despite the now accessible investigation of phase transitions into the very high pressure regime, experimental investigations of prototype ferroic perovskites above 50 GPa remain still scarce, with some notable exceptions among ferroelectrics (e.g. KNbO$_3$ \cite{Pruzan2007}, PbTiO$_3$ \cite{Janolin2008}), relaxor ferroelectrics (PbZn$_{1/3}$Nb$_{2/3}$O$_3$ \cite{janolin2006}), ferroelastics (SrTiO$_3$ \cite{Guennou2010}). Explorations of pressure-temperature or pressure-substitution phase diagrams remain even rarer \cite{Kreisel2009}.

Lanthanum aluminate (LaAlO$_3$ -- LAO) is a prototype compound for soft-mode driven antiferrodistorsive phase transitions. At ambient conditions, LAO crystallizes in the $R\overline 3c$ space group, as a result of the condensation of a soft mode at the $R$ point of the Brillouin zone boundary. At ambient conditions, the rhombohedral cell has the lattice parameters $a_R = 5.357$~\AA{} and $\alpha_R=60.12^{\circ}$, or in the hexagonal setting $a_H = 5.366$~\AA{} and $c_H = 13.109$~\AA{} \cite{Zhao2004a}. In this structure, the oxygen octahedra are rotated along the $[111]_C$ direction of the parent cubic cell. The rhombohedral distortion can be described by this single tilt angle. Following early structural studies \cite{Geller1956,Scott1969}, the temperature-induced phase transition at 813~K has been studied in great details (e.g. \cite{Hayward2005} and references within). On the theoretical side, the parameters of a Landau potential were fitted to available experimental data \cite{Carpenter2010b}. The pressure-induced rhombohedral to cubic transition of LAO was revealed by a powder Raman spectroscopy and synchrotron diffraction study \cite{Bouvier2002}. It was at that time the first exception to the so-far general rule stating that tilt angles in antiferrodistorsive perovskites should increase under pressure \cite{Samara1975}. This example has motivated theoretical work to explain this behaviour and formulate new rules and models to predict qualitatively and quantitavely the evolution of distortions (tilt angles) under pressure \cite{Zhao2004,Angel2005,Tohei2005,Zhao2006}. The evolution of the tilt angles in LAO itself was investigated by single crystal diffraction under hydrostatic stress up to 8 GPa \cite{Zhao2004a} and non-hydrostatic stress \cite{Zhao2011}.

However, several aspects of the high-pressure phase transition remain unclear. From their single-crystal diffraction study, Zhao \etal{}\cite{Zhao2004a} have extrapolated a transition pressure of 17.6~GPa, significantly higher than the value found by powder diffraction in \cite{Bouvier2002}. The reason for this difference remains unclear. Moreover, the ab-initio calculations in \cite{Luo2008} have predicted a volume drop at the pressure-induced transition of $\Delta V = 1$~\% at 15.4 GPa which was not seen in the powder diffraction study and requires experimental confirmation. In the light of these previous studies, the purpose of this high-pressure diffraction experiment was (i) to determine the transition pressure on a single crystal, (ii) to check the evolution of the volume and the order of the transition and finally (iii) to extend the structural study and search for further phase transitions up to the so far unreached 60~GPa pressure range.

\section{Experimental methods}

LAO single crystals were high-quality $[001]_C$-oriented substrates purchased from CrysTech GmbH and polished to a thickness of less than 10 $\upmu$m. The crystal selected had a lateral extension of 20~$\upmu$m and exhibited at room conditions micron-sized domains clearly visible under polarized light. The experiment was performed in diamond-anvil cells (DAC). The diamonds have the Boehler-Almax design with a cullet of 250~$\upmu$m. The pressure chamber was sealed by a rhenium gasket pre-indented to a thickness of about 40~$\upmu$m. Helium was used as pressure-transmitting medium. A small ruby sphere was loaded into the cell for pressure measurement using the classical fluorescence method. The pressure scale used was the equation given by Jacobsen \cite{Jacobsen2008}. This equation assumes the general form of the classical pressure scale by Mao \cite{Mao1986} where the coefficients have been reevaluated from the equation of state of MgO measured in hydrostatic conditions (in helium). Although the use of either of these pressure scale is indifferent within the experimental uncertainties for moderate pressures (say less than 1~\% between 0 and 10 GPa), the Mao equation underestimates the pressure by 3 GPa at 60 GPa \cite{Jacobsen2008}. In our experiment, the uncertainty on the pressure is taken as the pressure difference before and after the collection of diffraction patterns.

The x-ray diffraction experiment was performed on the ID27 beamline at the ESRF. The beam was monochromatic with a wavelength of 0.3738~\AA\ selected by an iodine K-edge filter and focused to a beam size of about 3~$\upmu$m. The patterns were collected in the rotating crystal geometry on a CCD detector with $-29^\circ\le\omega\le 29^\circ$ in $1^\circ$ steps. A precise calibration of the detector parameters was performed with a reference LaB$_6$ powder. The diffraction patterns from single crystal measurements were indexed with a home-made program based on the Fit2D software \cite{Hammersley1996}. The refinement of the lattice constants from the peak positions (50 to 70 reflections for each run) was performed with the program UnitCell \cite{UnitCell}. The fitting of equations of state was performed with EoSFit 5.2 \cite{eosfit}.

\section{Results}

In the following, the reflection indices for the cubic or pseudo-cubic cell $(hkl)_C$ are given for a cell doubled in the three directions with respect to the parent cubic cell. 

\subsection{Rhombohedral to cubic phase transition}

The evolution of the rhombohedral distortion can in principle be followed from the evolution of the lattice constants $a_{\mathrm H}$ and $c_{\mathrm H}$. As the crystal goes from rhombohedral to cubic, the ratio $c_{\mathrm H}/(a_{\mathrm H}\sqrt{6})$ goes to 1. However, the very small rhombohedral distortion ($\Delta d/d\approx 2.10^{-3}$ at ambient conditions) was below the resolution of our experiment, and the peak splittings expected for our polydomain crystal could not be observed. The comparison of peak widths, e.g. the doublet $(640)_{\mathrm C}\rightarrow(1\,2\,10)_{\mathrm H}+(232)_{\mathrm H}$ vs. the singlet $(600)_{\mathrm C}\rightarrow(036)_{\mathrm H}$, does show a line narrowing due to the evolution toward a cubic structure upon increasing pressure (not shown here), but this does not allow the calculation of the rhombohedral lattice constants, and remains a poor way to determine a transition pressure. Therefore, we only determined pseudo-cubic lattice constants in the rhombohedral phase (given in table \ref{tab:latticeconstants}), and turned to the observation of superstructure reflections for the study of the phase transition.   

\begin{figure}[tb]
\begin{center}
\includegraphics[width=0.48\textwidth]{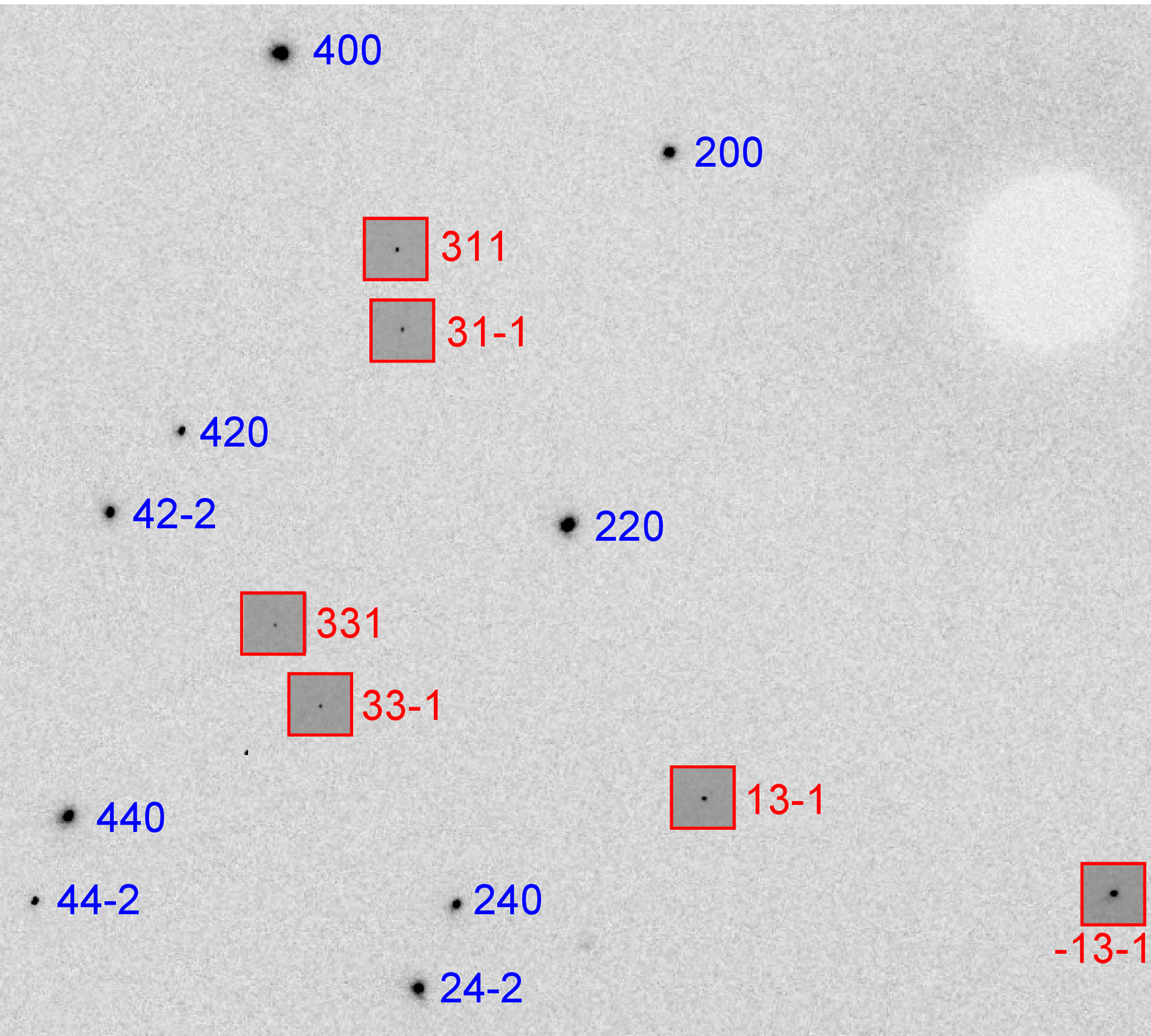}
\caption{Extract of a diffraction pattern. The indices are given for the doubled cubic unit cell. The superstructure reflections (in red), barely visible, are taken here from an overexposed pattern for clarity.}
\label{fig:cliche}
\end{center}
\end{figure}

In distorted perovksites, superstructure reflections arise from tilts of the octahedra (or antiferroelectric cation shifts). Following Glazer's notations \cite{Glazer1972,Glazer1975,Mitchell2002}, the tilt system in LAO is noted $a^-a^-a^-$ and gives rise to superstructure reflections that can be indexed in the doubled cubic cell with the general form $(hkl)_C$ where $h$, $k$ and $l$ are odd integers and one index at least is different from the others. Note that for this particular tilt system, the presence of a domain structure is indifferent for the study of the superstructure intensities: all domains give rise to superstructure reflections at the same spots in reciprocal space, within our experimental resolution. The superstructure intensity therefore sums up contributions from all the domains present in the crystal and is not affected by the domain structure, nor a change in domain volume ratio. The intensity of the superstructure scales like the square of the tilt angle; they remain very weak, of the order of 1/1000 compared to the intensity of principal Bragg peaks. An example of a diffraction pattern is given in figure \ref{fig:cliche} for illustration.

\begin{figure}[tb]
\begin{center}
\includegraphics[width=0.48\textwidth]{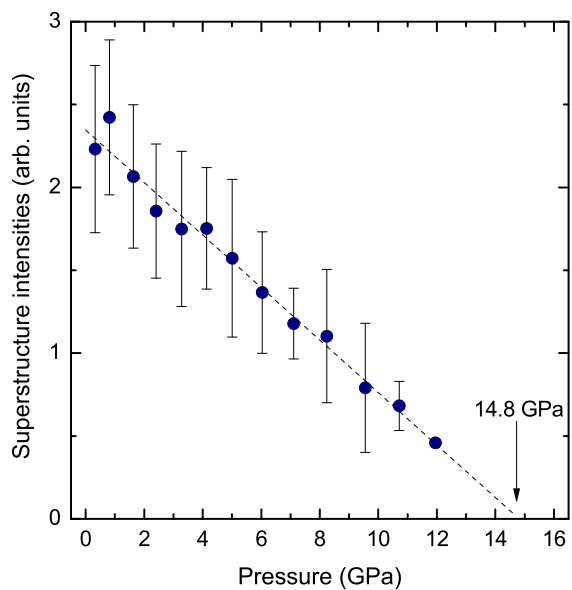}
\caption{Evolution of normalized superstructure intensities $(311)_C$ with increasing pressure. The values are calculated as the average of the intensities of all observed reflections and the error bars reflect their dispersion. For the last point at 12 GPa, only one reflection could be measured.}
\label{fig:transition}
\end{center}
\end{figure}

In our diffraction patterns, all expected sets of superstructures $(311)_C$, $(331)_C$, $(531)_C$ etc. were observed. However, only the reflections of the first set were intense enough to be followed with reasonable accuracy. Their average intensities, normalized with respect to the intensity of neighbouring Bragg peaks are given in figure \ref{fig:transition}. The intensity decreases linearly and cancels out at a critical pressure $P_c=14.8(4)$~GPa when extrapolated to higher pressure. This transition pressure is consistent with the previous powder study \cite{Bouvier2002}, as well as the value calculated by ab-initio \cite{Luo2008} (15.4 GPa). There remains however a significant difference with the 17.6~GPa extrapolated from the single crystal data given in \cite{Zhao2004a}. This is likely due to the uncertainty in the extrapolation to the transition pressure, their experimental data ranging only up to 8~GPa.

After the transition, the crystal remains cubic up to the highest pressure investigated (63 GPa). This deserves a comment in the framework of the general rules for predicting the evolution of distortion of perovksites under pressure \cite{Zhao2004,Angel2005}. According to these rules, the change in tilt angles under pressure is determined by the compressibility ratio of the $A$O$_{12}$ and $B$O$_6$ polyhedra, which are the building blocks of the perovskite structure. It was found that for most $A^{3+}B^{3+}$O$_3$, such as rare-earth aluminates, the tilt angles should decrease under pressure, or, stated in terms of vibrational properties, that associated phonon modes should soften. This implicitely suggests that a crystal such as LAO should remain cubic after its transition. However, the validity limit of this assumption, or whether a new distortion might occur at much higher pressure has not been verified or discussed. Our experiment shows that the assumption of a cubic structure remains valid over a pressure range of 40 GPa for a compression $V/V(P_c)$ of 86~\%, which is, to the best of our knowledge, the widest compression range over which this has been verified in perovskites.

\subsection{Compressibility and equations of state}

We focus first on the equation of state for the cubic phase. We used a third-order Birch-Murnaghan equation of state, with the transition pressure $P_c=14.8$~GPa as a reference pressure. This calculation yielded $K(P_c)=265.3(2.6)$~GPa, $K'(P_c)=3.4(1)$ and $V(P_c)=51.163(13)$~\AA$^3$ for a weighted $\chi^2=0.35$. In the rhombohedral phase, a second-order Birch-Murnaghan equation of state was found to model adequately the experimental data. Choosing again the transition pressure as a reference, we find $V(P_c)=51.189(15)$~\AA$^3$ and $K(P_c)=255.2(1.4)$~GPa, $K'(P_c)=4$ being constrained by the equation. The differences between measured and calculated pressures remain within experimental uncertainties for both refinements. In order to compare our data with previous studies, we have also calculated the parameters $V_0$, $K_0$ and $K'_0$ for both the low-pressure and the high-pressure phases. This can be done either by extrapolating our equations of state, or making new refinements to the data with 1~bar as a reference pressure. The two procedures are not equivalent but different tests have shown that both methods yield in that case similar results given our experimental uncertainties. We report the outcome of the second method together with data from the litterature in table \ref{tab:eos}.

\begin{figure}[tb]
\begin{center}
\includegraphics[width=0.48\textwidth]{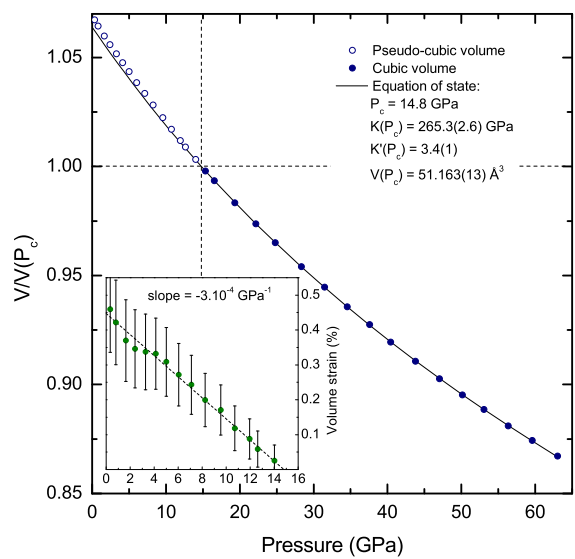}
\caption{Evolution of the relative volume $V/V(P_c)$. In the rhombohedral phase only a pseudo-cubic volume could be determined. The line is a third-order Birch-Murnaghan equation of state with parameters fitted from the data in the cubic phase. (Inset) Pseudo-volume strain in the rhombohedral phase calculated as $V/V_{\mathrm{EoS}}-1$, where $V$ is the experimental pseudo-cubic volume and $V_{\mathrm{EoS}}$ is the volume given by the extrapolation of the equation of state for the cubic phase in the stability region of the rhombohedral phase.}
\label{fig:volume}
\end{center}
\end{figure}

The evolution of the volume predicted by the equation of state in the cubic phase is given in figure \ref{fig:volume} and compared to the experimental values. In the rhombohedral phase, the experimental volume clearly deviates from the equation of state, reflecting the presence of a volume strain in the low-symmetry phase, which can be calculated by $e_a = V/V_{\mathrm{EoS}}-1$, $V_{\mathrm{EoS}}$ being the volume calculated from the cubic equation of state and extrapolated down to ambient pressure, bearing in mind that the experimental volume is a pseudo-cubic volume only. This pseudo volume strain is plotted in figure \ref{fig:volume} (inset). The uncertainty is intrinsically large, as it results from uncertainties in both the experimental volume and the extrapolation of the equation of state. Still, it shows the linear pressure dependence expected for a second-order phase transition with a slope of -3.10$^{-4}$~GPa$^{-1}$. 

\begin{figure}[tb]
\begin{center}
\includegraphics[width=0.48\textwidth]{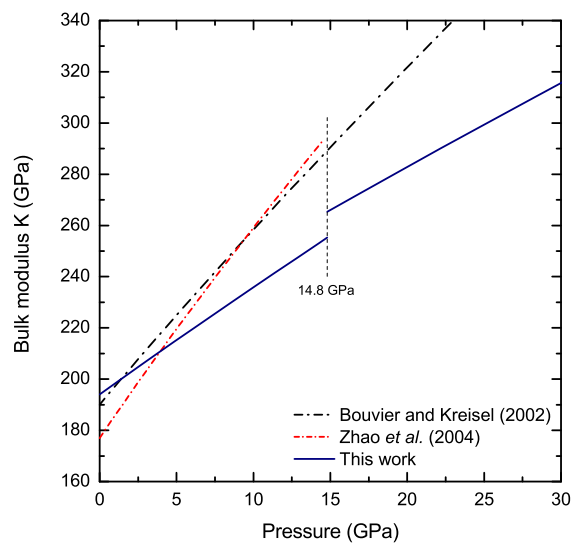}
\caption{Pressure-dependance of the bulk modulus $K$ as predicted by the equations of state proposed by Bouvier and Kreisel \cite{Bouvier2002}, Zhao \etal{} \cite{Zhao2004a} and in this work.}
\label{fig:Keos}
\end{center}
\end{figure}

The comparison with previous studies calls for the following comments. First, we do not confirm the 1~\% volume jump at the transition predicted in \cite{Luo2008}. Instead, the smooth evolution of the volume is consistent with the second-order character of the transition. Second, the jump of the bulk modulus at the transition is calculated as $\Delta K=10(4)$~GPa. This is of the order of magnitude of the value predicted by the Landau model proposed in \cite{Carpenter2010a}. Finally, we want to examine the bulk moduli in more details. The agreement between the different values of $K_0$ is very good, including the theoretical calculations in \cite{Luo2008} (table \ref{tab:eos}). However, the values of $K'_0$ determined from the high-pressure diffraction experiments differ significantly, which has important implications for the pressure evolution of $K$. In figure \ref{fig:Keos}, we show the evolution of the bulk modulus as predicted by the different models. The two equations of state given in \cite{Bouvier2002,Zhao2004a} are seemingly in good agreement. However, one must bear in mind that the equation of state in ref. \cite{Bouvier2002} was determined from pseudo-cubic volume values whereas in ref. \cite{Zhao2004a}, the equation of state is derived from volume values measured in the rhombohedral phase only. As a result, the two equations of state cannot be compared, and the apparent agreement at low pressures must be regarded as coincidental. On the other hand, there is an important difference between our model and the model determined from the powder diffraction data in \cite{Bouvier2002}, which can most probably be accounted for by the different pressure-transmitting media used (N$_2$ in \cite{Bouvier2002} and He in this work). The experiment reported in \cite{Zhao2004a} on the contrary was performed in the usual 4:1 methanol-ethanol mixture which, like helium, remains liquid in the 0--8 GPa pressure range and as such provides very good hydrostatic conditions. Their equation of state for the rhomboedral phase together with our equation of state for the cubic phase should therefore provide a good description of the bulk modulus over the full pressure range 0-63~GPa. However, the inspection of the data shows that it is not the case: due to the unusually large value of $K'_0=8.9$, the bulk modulus predicted by the equation of state from ref. \cite{Zhao2004a} increases very rapidly at low pressures and reaches 295~GPa at the transition. This is inconsistent with our equation of state in the cubic phase, as it would lead to a thermodynamically problematic negative $\Delta K$ at the transition. We have checked that different choices of ruby pressure scales have only a marginal influence on the values and do not solve this issue. The explanation thus remains an open question.

\section{Conclusion}

We have described a high-pressure synchrotron single-crystal diffraction on a LaAlO$_3$ single crystal. We have confirmed the transition from rhombohedral to cubic at $P_c=14.8$~GPa through the observation of the superstructure intensities, whose linear pressure dependence confirmed the second-order character of this transition. Remarkably, after the transition, the cubic structure is stable up to 63 GPa. Last, we have determined the parameters of a third-order Birch-Murnaghan equation of state for the cubic phase: $K(P_c)=265.3(2.6)$~GPa, $K'(P_c)=3.4(1)$ and $V(P_c)=51.163(13)$~\AA$^3$. However, open questions regarding the compressibility in the rhombohedral phase remain.

\section*{Acknowledgments}

We are grateful to the ESRF staff, especially M. Mezouar for allocation of inhouse beamtime. Support from the French National Research Agency (ANR Blanc PROPER) is acknowledged.

\section*{References}

\bibliographystyle{unsrt}
\bibliography{biblio}

\clearpage

\begin{table}[tb]
\begin{center}
\caption{Lattice parameters as a function of pressure. In the rhombohedral phase, the pseudo-cubic lattice constant results directly from a fit of the peak positions in the cubic system.}
\label{tab:latticeconstants}
\begin{tabular}{@{}c c c c c@{}}
\hline\hline
\multicolumn{2}{c}{Rhombohedral phase} && \multicolumn{2}{c}{Cubic phase} \\
\cline{1-2} \cline{4-5}
$P$ (GPa) 	& $a_{\mathrm{pc}}$ (\AA) && $P$ (GPa) 	& $a$ (\AA) \\
0.33(1)$\phantom 0$ 		& 3.7938(2) 	& & 14.02(13) 	& 3.7164(2) \\
0.81(5)$\phantom 0$ 		& 3.7905(2) 	& & 15.35(10) 	& 3.7099(2) \\
1.63(10) 	& 3.7851(2) 	& & 16.55(13) 	& 3.7043(2) \\
2.41(10) 	& 3.7803(2) 	& & 19.32(18) 	& 3.6913(2) \\
3.28(15) 	& 3.7753(2) 	& & 22.16(4)$\phantom 0$ 	& 3.6796(2) \\
4.14(9)$\phantom 0$ 		& 3.7705(2) 	& & 24.81(14) 	& 3.6686(2) \\
5.00(13) 	& 3.7655(2) 	& & 28.34(10) 	& 3.6546(2) \\
6.04(12) 	& 3.7594(2) 	& & 31.47(13) 	& 3.6423(2) \\
7.11(15) 	& 3.7534(2) 	& & 34.53(12) 	& 3.6310(2) \\
8.24(14) 	& 3.7470(2) 	& & 37.57(1)$\phantom 0$ 	& 3.6204(2) \\
9.55(11) 	& 3.7400(2) 	& & 40.42(10) 	& 3.6099(2) \\
10.72(17) 	& 3.7335(2) 	& & 43.80(12) 	& 3.5984(2) \\
11.95(12) 	& 3.7270(2) 	& & 47.04(11) 	& 3.5877(2) \\
12.63(16) 	& 3.7234(2) 	& & 50.13(9)$\phantom 0$ 	& 3.5779(2) \\
 				& 					& & 53.05(10) 	& 3.5690(2) \\
 				& 					& & 56.36(2)$\phantom 0$ 	& 3.5589(2) \\
 				& 					& & 59.58(9)$\phantom 0$ 	& 3.5498(3) \\
 				& 					& & 63.05(20)	& 3.5401(3) \\
\hline\hline
\end{tabular}
\end{center}
\end{table}

\begin{table}[tb]
\begin{center}
\caption{Parameters of the equation of state from this work compared to the values determined in previous studies. For the ab-initio calculations in \cite{Luo2008}, values obtained by the local density approximation (LDA) and the generalized gradient approximation (GGA) are both given.}
\label{tab:eos}
\begin{tabular}{@{}l l m{1.8cm} m{1.8cm} c m{1.8cm} m{1.8cm}@{}}
\hline\hline
& & \multicolumn{2}{c}{Cubic phase} & & \multicolumn{2}{c}{Rhombohedral phase}\\
\cline{3-4}\cline{6-7}
									& 				& $K_0$ (GPa)	& $K'_0$  	& & $K_0$ (GPa)& $K'_0$ 	\\  
Ref. \cite{Luo2008} 			& LDA			& 188				& 				& & 196 			& 				\\
									& GGA			& 199 			& 				& & 186 			& 				\\
This work 						& XRD			& 215(4)			& 3.6(1) 	& & 196(2)		& 4			\\ 
Ref. \cite{Bouvier2002} 	& XRD 		& 					& 				& & 190(5)		& 7.2(4)		\\
Ref. \cite{Zhao2004a} 		& XRD 		& 					& 				& & 177(4)		& 8.9(1.6)	\\
Ref. \cite{Carpenter2010a}	& Brillouin & 					& 				& & 196			& 				\\
\hline\hline
\end{tabular}
\end{center}
\end{table}

\end{document}